\documentclass[prb,titlepage,superscriptaddress,reprint,nobibnotes]{revtex4-1}

\usepackage{amsmath, amssymb, amsfonts,mathrsfs}
\usepackage[english]{babel}
\usepackage{natbib}
\usepackage{graphicx}
\usepackage{xcolor}

\newcommand{\ket}[1]{\left\vert{#1}\right\rangle}

\newcommand{\romaUNI}{Dipartimento di Fisica - Sapienza Universit\`{a} di Roma, P.le Aldo Moro 5, I-00185 Roma (RM), Italy}
\newcommand{\milanoCNR}{Istituto di Fotonica e Nanotecnologie - Consiglio Nazionale delle Ricerche (IFN-CNR), P.za Leonardo da Vinci, 32, I-20133 Milano (MI), Italy}
\newcommand{\milanoPOLI}{Dipartimento di Fisica - Politecnico di Milano, P.za Leonardo da Vinci, 32, I-20133 Milano (MI), Italy}
\newcommand{\parigiTELECOM}{(present address) T\'el\'ecom ParisTech, CNRS-LTCI, 46 rue Barrault, F-75634 Paris CEDEX 13, France}
\newcommand{\bristol}{(present address) Centre for Quantum Photonics H. H. Wills Physics Laboratory and Department of Electrical and Electronic Engineering, University of Bristol, Merchant Ventures Building, Woodland Road, Bristol BS8 1 UB, UK.}

\begin{document}

\title[Hyperentangled photon states on a chip]{Path-polarization hyperentangled and cluster states of photons on a chip \\ Running Title: Hyperentangled photon states on a chip}

\author{Mario Arnolfo Ciampini}
\affiliation{\romaUNI}
\author{Adeline Orieux}
\affiliation{\romaUNI}
\affiliation{\parigiTELECOM}
\author{Stefano Paesani}
\affiliation{\romaUNI}
\affiliation{\bristol}
\author{Fabio Sciarrino}
\affiliation{\romaUNI}
\author{Giacomo Corrielli}
\author{Andrea Crespi}
\author{Roberta Ramponi}
\author{Roberto Osellame}
\affiliation{\milanoCNR}
\affiliation{\milanoPOLI}
\author{Paolo Mataloni}
\affiliation{\romaUNI}
\email{paolo.mataloni@uniroma1.it}



\begin{abstract}

Encoding many qubits in different degrees of freedom (DOFs) of single photons is one of the routes towards enlarging the Hilbert space spanned by a photonic quantum state. Hyperentangled photon  states (i.e. states showing entanglement in multiple DOFs) have demonstrated significant implications for both fundamental physics tests and quantum communication and computation. Increasing the number of qubits of photonic experiments requires miniaturization and integration of the basic elements and functions to guarantee the set-up stability. This motivates the development of technologies allowing the precise control of different photonic DOFs on a chip. We demonstrate the contextual use of path and polarization qubits propagating within an integrated quantum circuit. We tested the properties of four-qubit linear cluster states built on both DOFs and we exploited them to perform the Grover's search algorithm according to the one-way quantum computation model. Our results pave the way towards the full integration on a chip of hybrid multiqubit multiphoton states.

\end{abstract}


\maketitle

\section{Introduction}

Novel integrated photonic circuits built on a single chip have been recently introduced within the realm of quantum information \cite{Politi08}, disclosing new perspectives towards quantum communication \cite{Barreiro08}, quantum computation \cite{Politi09}, and the quantum simulation of physical phenomena \cite{Feynman82,Lloyd96,Crespi13,Spring13,Tillmann13,Broome13}. The next generation of integrated quantum circuits (IQCs), incorporating highly efficient photon sources \cite{Kruse15,Herrmann13,Silverstone14, Silverstone15} and detectors \cite{Goltsman01,Gaggero10,Sprengers11,Pernice12,Sahin13}, are expected to have a large impact in future photonic quantum technologies and are essential to achieve a level of complexity and stability of the operations higher than what previously demonstrated. The miniaturization of integrated photonic devices represents a necessary step towards the implementation of state-of-the art quantum information protocols, which require an exponentially high number of elements and an increasing stability, impossibile to achieve with standard bulk optical setup.
Inherently stable interferometer networks, composed of waveguides, beam splitters and phase shifters, built in two dimensions on different material platforms, such as silicon, silicon nitride and others, are realized by lithography, a well established technique already developed for telecom wavelengths. This approach allows the fabrication of a large number of replicas of the same circuit by using a single mask and represents at the moment the strongest candidate for a large-scale production of IQCs. While it has been demonstrated that the operation complexity performed by such systems may be very high, it is worth to remember that they are still limited by the need of using a large number of photons. Moreover, only path-encoded qubits are allowed in such systems, since polarization qubits are degraded by the intrinsic large birefringence of the material substrate or of the waveguide itself. On the other hand, several applications in the quantum domain, such as quantum computation and quantum communications, may greatly benefit from the possibility of manipulating and controlling polarization qubits.

Femtosecond laser writing, recently introduced for IQC applications \cite{Marshall09,Sansoni10,Heilmann2014}, allows to write in three dimensions circular transverse waveguide profiles able to support the propagation of nearly Gaussian modes with any polarization state, while keeping highly stable the phase of path-encoded qubits. Besides, this technique makes it possible to perform arbitrary transformations of the polarization state by suitable integrated devices, such as polarization beam splitters \cite{Crespi11} and  waveguide-based optical waveplates \cite{Heilmann2014, Corrielli14}.

Nowadays, the building blocks necessary to perform the basic operations with path- and polarization-qubits are available. Time is right to demonstrate the simultaneous control of two different degrees of freedom of the photons within the same chip. In this work we manipulate path- and polarization-encoded qubits belonging to a 2-photon 4-qubit hyperentangled/cluster state \cite{Barbieri05,Cinelli05,Barreiro05,Vallone07,Chen07} and propagating through an integrated quantum circuit.
This enabled us to demonstrate the Grover's algorithm on a four element database, in a one-way quantum computing approach.

\section{Materials and Methods}
\label{sec:examples}
\begin{figure*}[tbh]
\centering
\includegraphics{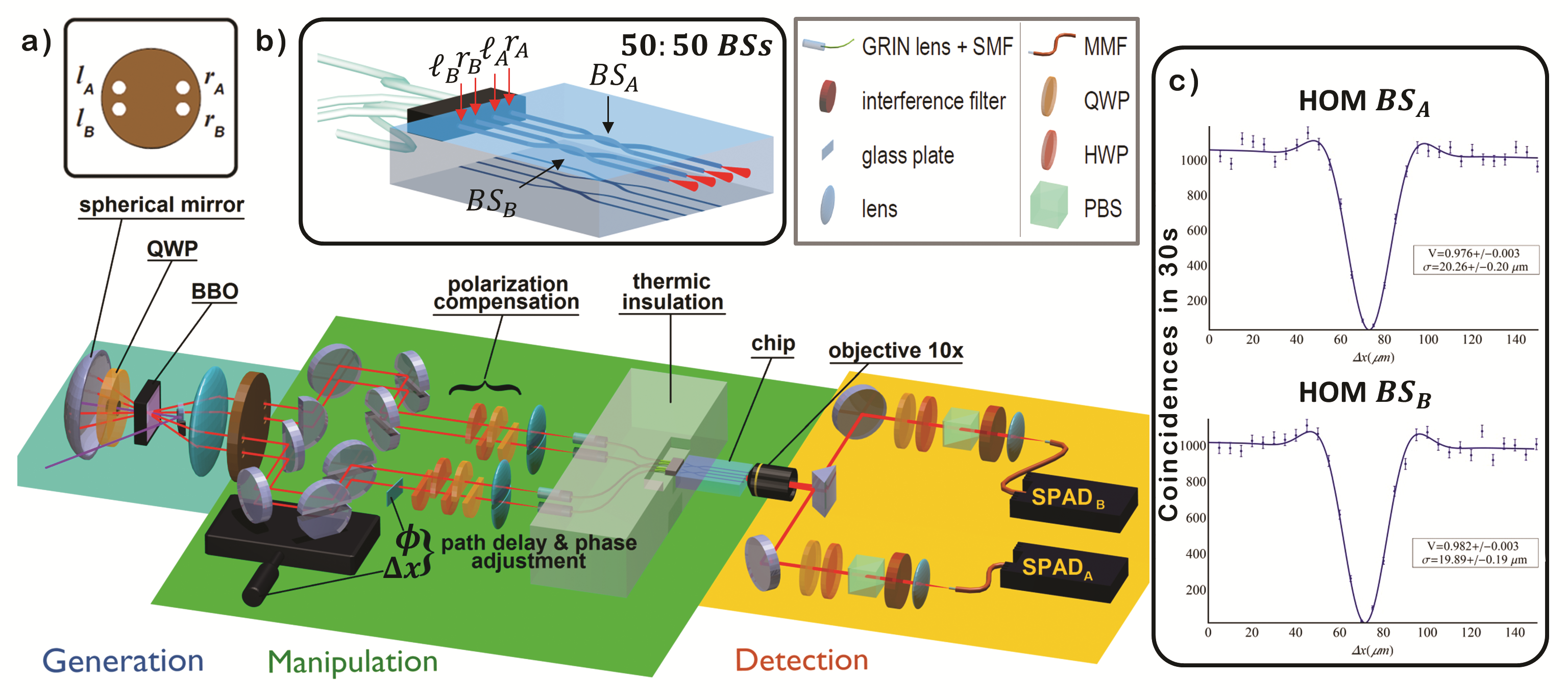}
\caption{\textbf{Experimental setup.} The state $|\Omega\rangle$ is generated by the hyperentangled source in blue area. The 4-hole screen in \textbf{a)} selects the spatial modes $\ell_A$, $\ell_B$, $r_A$, $r_B$. In the green area the four modes are independently addressed using half mirrors, half wave plates and half lenses so to compensate their polarization and to couple them into the fiber array. A translation stage is used to control path indistinguishability between left and right modes, a glass plate in mode $\ell_B$ is used to control the path phase. The chip is connected to the fiber array using a NanoMax 6-axis stage. In the orange area measurements are performed: two of the outputs of the chip are addressed into multi-mode fibers connected to SPADs in coincidence mode. Polarization analysis is performed using QWP, HWP and PBS. \textbf{b)} Schematic representation of the integrated device. \textbf{c)} Hong-Ou-Mandel dip of $BS_A$ and $BS_B$. }
\label{fig1}
\end{figure*}

\begin{figure}[tbh]
\centering
\includegraphics{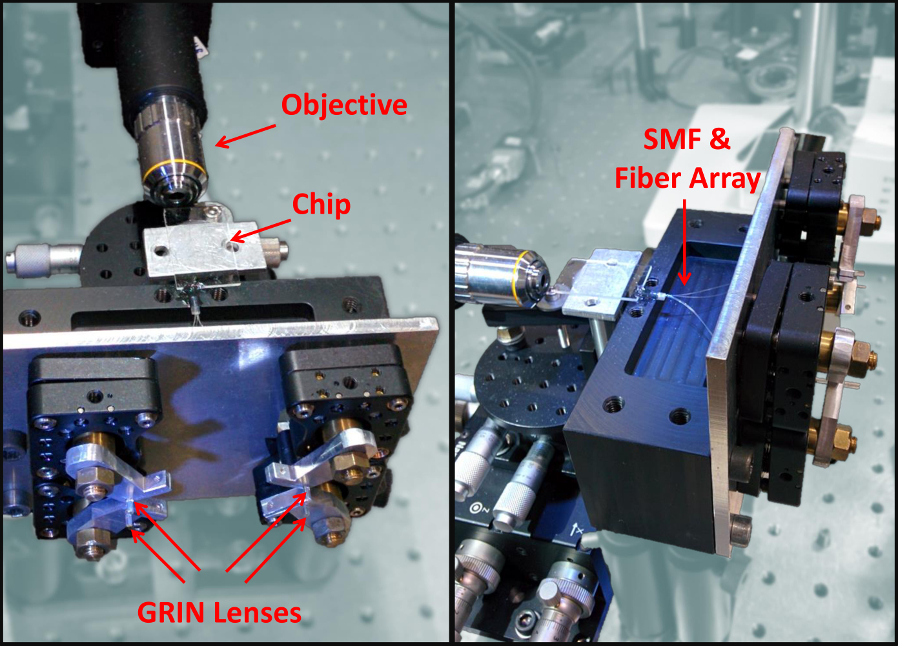}
\caption{\textbf{Chip mount assembly.}}
\label{figexp}
\end{figure}

Our experimental setup is depicted in Fig.~\ref{fig1} and Fig.~\ref{figexp}   and consists of a hyperentangled-photon source, a manipulation stage, which includes the integrated photonic chip, and a detection stage.
The source generates pairs of photons, hyperentangled in the path and polarization DOFs, via spontaneous-parametric down conversion at 710~nm by a BBO type I nonlinear crystal \cite{Barbieri05}. Polarization entanglement is produced by optical superposition of two cones of perpendicular polarization created by double passage, back and forth, of a UV laser pump through the BBO crystal from both sides \cite{Cinelli04}. Path entanglement is generated by selecting with a 4-hole screen two pairs of correlated spatial modes, namely $\ell_A$, $r_B$ and $r_A$, $\ell_B$.

The generated hyperentangled (HE) state is:

\begin{equation}
\ket{\Omega} =\frac{1}{\sqrt{2}} ( \ket{H_A H_B}+e^{i\theta} \ket{V_A V_B}) \otimes \frac{1}{\sqrt{2}} ( \ket{r_A \ell_B} + e^{i\phi} \ket{\ell_A r_B} ),
\label{eq:he}
\end{equation}
where $A$ and $B$ label the two photons; $r_A$, $\ell_A$, $r_B$, and $\ell_B$ identify the four spatial modes, while $H$ and $V$ represent the two possible polarization states for each photon. As will be discussed in the following, the phases $\theta$ and $\phi$, corresponding respectively to the polarization and path DOFs, can be independently controlled.

The chip consists of two waveguide balanced beam splitters ($BS_A$ and $BS_B$), yielding polarization insensitive behaviour\cite{Sansoni12}, fabricated by femtosecond laser  waveguide writing\cite{DellaValle09} using the second harmonic ($\lambda$=515~nm) of a  Yb:KYW cavity-dumped laser oscillator (300~fs pulse duration, 1~MHz repetition rate). Femtosecond laser pulses are focused by a 0.6~NA microscope objective into the volume of the glass substrate (EAGLE 2000, Corning), where nonlinear energy absorption creates a permanent and localized refractive index increase. Waveguides are produced by smoothly translating the sample under the laser beam, using Aerotech FiberGLIDE 3D air-bearing stages. Under proper irradiation conditions (100~nJ pulse energy and 10~mm~s$^{-1}$ translation speed) single-mode waveguides at 710~nm are produced, at 170~$\mu$ depth below the glass surface, characterized by a mode diameter of $\sim$8~$\mu$m, propagation loss of 1.5~dB~cm$^{-1}$ and coupling loss to single mode fibers $<$~1~dB per facet. Integrated beam splitters are realized following a particular three-dimensional directional-coupler design\cite{Sansoni12} that ensures insensitivity to the polarization. To achieve a balanced splitting ratio the waveguides, initially 127~$\mu$m far, are brought closer (with a bending radius of 90~mm) down to 10~$\mu m$ distance for an interaction length of 1.8~mm. Overall chip length is 34~mm.

The four modes $r_A$, $\ell_A$ and $r_B$, $\ell_B$ are coupled to the input ports of $BS_A$ and $BS_B$, respectively, through a 8 cm long, single mode fiber array, terminated at the input side by a set of four Gradient Index (GRIN) lenses. One of the main technical 
issue is given by the independent injection of each mode into the corresponding GRIN lens; this is performed by using for each mode a set of custom made half-mirrors and one half-lens. Besides, polarization compensation is individually performed for each mode through a properly chosen set of half waveplates (HWPs) and quarter waveplates (QWPs). Finally, the fiber array is thermally insulated from the environment in order to guarantee path stability.
Output light from the chip is collected by a 10$\times$ objective. The coupling ratios of the 4 modes in each optical component and the overall transmission efficiency are provided in the Supplementary Information.
Two interference filters centred at 710~nm select 10~nm bandwidth and ensure photon indistinguishability, and two avalanche photodiode (APD) detectors measure coincidences between output modes $r'_A$, $\ell'_B$ that are coupled to the detectors through multi mode fibers (we label $\ell'_A$, $r'_A$  and $\ell'_B$, $r'_B$ the output modes of $BS_A$ and $BS_B$ respectively).

Figure~\ref{fig1}c shows the Hong-Ou-Mandel (HOM) dips obtained when the two photons are injected within $BS_A$ and $BS_B$, respectively. We obtained the following visibilities: $\mathcal{V}_A=(0.976\pm 0.003)$ for $BS_A$ and $\mathcal{V}_B=(0.982\pm 0.003)$ for $BS_B$, thus showing the correct operation of the two systems. After removing $N_{acc}=12$ accidental coincidences every 30 sec, we obtain $\mathcal{V}_{A-net}=(0.985\pm 0.003)$ and $\mathcal{V}_{B-net}=(0.991\pm 0.003)$.

\section{Results and Discussion}

\begin{figure}[bt]
\centering
\includegraphics{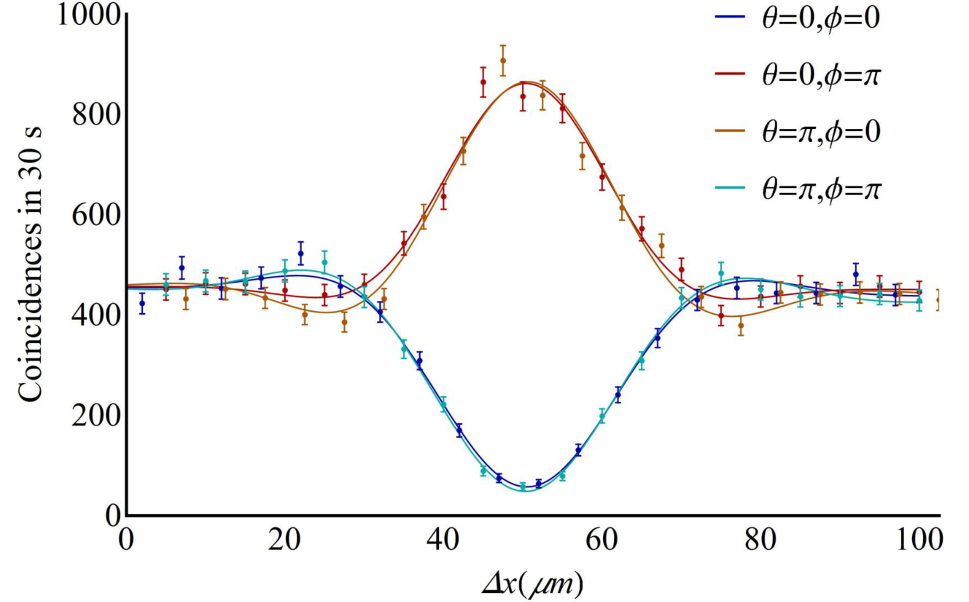}
\caption{\textbf{Hyperentangled state interference.} Interference pattern between modes $|\ell_A r_B\rangle$ and $|r_A \ell_B\rangle$ varying both $\theta$ and $\phi$. Peaks and dips derive from the symmetry of the entire HE wave function. Experimental data of each curve are fitted with the convolution of a gaussian with a sinc function; error bars derive from the Poissonian statistical distribution of counts. Errors on the peak/dip visibilities are estimated by using a Monte Carlo statistical analysis on experimental data.}
\label{fig:he01}
\end{figure}

In a first experiment, we injected in the chip the path-polarization HE state:
\begin{equation}
|\Xi \rangle = |\Psi^\pm_\pi \rangle |\Theta^\pm_k\rangle,
\label{eq:xi}
\end{equation}
where $|\Psi^\pm_\pi \rangle = \frac{1}{\sqrt{2}}(|H_A V_B\rangle \pm |V_A H_B\rangle)$ and $|\Theta^\pm_k \rangle = \frac{1}{\sqrt{2}}(|\ell_A r_B\rangle \pm |r_A \ell_B\rangle)$. This state was obtained from Eq.~\eqref{eq:he} by introducing the polarization transformation using the compensation waveplates shown in Fig.~\ref{fig1} such that $H \rightarrow H$ and $V \rightarrow V$ on modes $\ell_A$ and $r_A$ while $H \rightarrow V$ and $V \rightarrow H$ on $\ell_B$ and $r_B$. In this way, we were able to guarantee the polarization compensation over the entire system, which includes the fiber array and the chip, on the computational polarization basis. In order to compensate the two pairs of correlated modes on the diagonal basis, we introduced before the compensation plates on mode $\ell_B$  an additional half HWP at zero degrees. As the addition of the plate preserves the compensation in the computational basis, we tilted it along its vertical axis in order to keep constant the phase difference between the two couples of modes. This ensures the simultaneous compensation on the polarization degree of freedom. Arbitrary values for the parameters $\theta$ and $\phi$ in Eq.~\eqref{eq:he} can be set by translating the spherical mirror within the HE source and by tilting an additional glass plate on mode $\ell_B$, respectively. The plus signs appearing in Eq.~\eqref{eq:xi} were achieved by setting $\phi, \theta=0$ and the minus signs are obtained by setting $\phi, \theta=\pi$. The overall symmetry of the state determines the behaviour of the two photons: if the wavefunction of the HE state is symmetric they emerge from the same output port of the beam-splitters, corresponding to a coincidence dip, while the expected result in the case of an antisymmetric wavefunction is a coincidence peak. This behaviour can be analysed by recovering the dips and peaks of path entanglement varying both $\phi$ and $\theta$ so that a dip can be obtained with $(\phi,\theta)=(0,0),(\pi,\pi)$ and a peak with $(\phi,\theta)=(0,\pi),(\pi,0)$.
Results are shown in Fig.~\ref{fig:he01}. The average peak/dip visibilities are $\mathcal{V}_{peak}=0.93\pm 0.20$ and $\mathcal{V}_{dip}=0.860\pm0.005$. These results are comparable with those of Ref.~\cite{Barbieri05} and prove the achievement of path-polarization hyperentanglement on chip with good fidelity.

\begin{figure}[bt]
\centering
\includegraphics[width=0.48\textwidth]{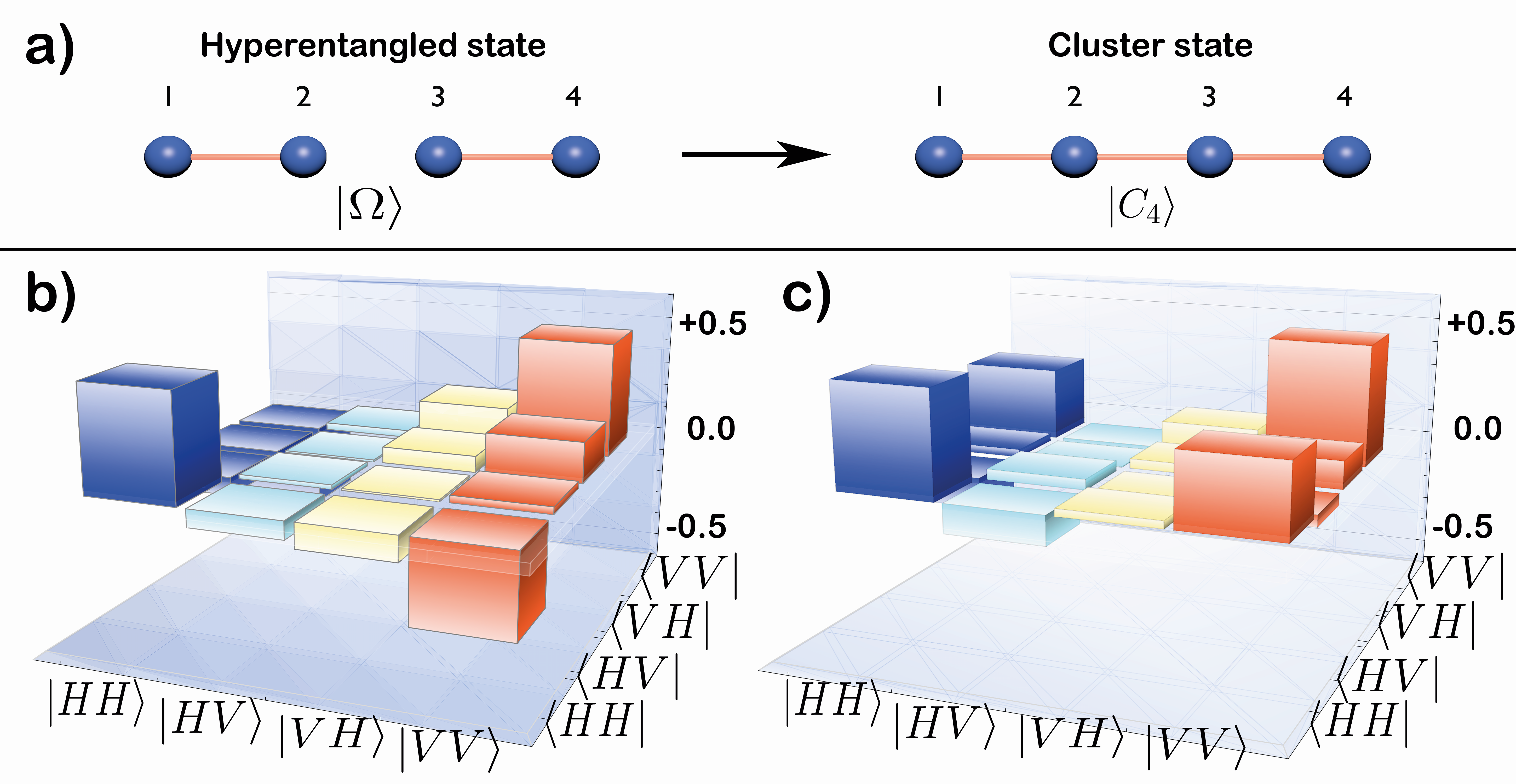}
\caption{\textbf{Cluster state.} \textbf{a)} Graphical representation of four qubits hyperentangled and cluster states.\textbf{ b)} Real part of the two qubits polarization tomography of pair $|\ell_A r_B\rangle$ of the cluster state, which is compensated to be a $|\Phi^-\rangle$ state.  \textbf{c)} Real part of the two qubits polarization tomography of pair $|r_A \ell_B\rangle$ of the cluster state, compensated to be a $|\Phi^+\rangle$ state. Imaginary parts of the two tomographies are negligible.
\label{fig:tomo_cluster}}
\end{figure}

In a second experiment, the same hyperentangled source was used to engineer a four-qubit cluster state encoded in the path and polarization DOFs of the two photons:
\begin{align*}
|C_4\rangle= \frac{1}{2}(|H_A r_A H_B \ell_B\rangle +|V_A r_A V_B \ell_B\rangle + |H_A \ell_A H_B r_B \rangle \\ - |V_A \ell_A V_B r_B \rangle) =  \frac{1}{\sqrt{2}}(|\Phi^+\rangle|r_A \ell_B\rangle + |\Phi^-\rangle |\ell_A r_B\rangle),
\label{eq:cluster}
\end{align*}
where $|\Phi^\pm \rangle = \frac{1}{\sqrt{2}}(|H_A H_B\rangle \pm |V_A V_B\rangle)$. At variance with standard hyperentangled states, four qubits cluster states are not biseparable and present genuine multipartite entanglement \cite{Briegel01}. They can be thought as graphs where the vertices are the physical qubits initially in the state $|+\rangle=(|0\rangle+|1\rangle)/\sqrt{2}$, and each edge represents a controlled-phase gate entangling the two connected nodes. As an example, in this graphical representation a HE state is equivalent, up to single qubit transformations, to the graph formed by two disjointed couples of vertices, as shown in Fig.~\ref{fig:tomo_cluster}a. A cluster state can then be obtained by connecting these pairs of qubits. In our case this was easily performed by inserting a zero-order HWP in mode $r_A$ oriented along the optical axis, as explained later. In the One-Way Quantum Computation model \cite{Raussendorf01}, cluster states provide the entire resource for the computation as the information is written, processed and read out by single-qubit measurements on the physical qubits of the cluster. This approach to quantum computation transfers the main complexity of the process from the ability to implement multi-qubit gates to the capability to create the initial cluster state. It is therefore well-suited for quantum optical schemes where states can be produced with high fidelity while photon-photon interactions are difficult to achieve, thus precluding two-qubits gates from linear optical circuits.

Starting from the HE state $|\Omega\rangle$ and inserting a zero order, zero degrees HWP in mode $r_A$ we were able to change the polarization of $|r_A \ell_B\rangle$ from $|\Phi^-\rangle$ to $|\Phi^+\rangle$ while keeping $|\ell_A r_B\rangle$ unchanged, thus creating the cluster state $|C_4\rangle$. The quantum state tomographies corresponding to the two mode pairs are reported in Fig. \ref{fig:tomo_cluster}b-c and correspond to the following parameters:
\begin{align*}
F_{\Phi^-}&=0.91\pm 0.10,  &C_{\Phi^-}&=0.88 \pm 0.08,\\
F_{\Phi^+}&=0.83 \pm 0.11, &  C_{\Phi^+}&=0.91\pm 0.08.
\end{align*}
Here $F$ is the fidelity and $C$ is the concurrence of the experimental state, while errors are calculated from a Monte Carlo analysis of the experimental data.
The obtained results are comparable with those of the tomographies of the hyperentanglement source reported in the SI, thus proving that the cluster state is correctly generated.
The stabilizer formalism, explained in Ref. \cite{Toth05} can be adopted to measure a genuine multipartite entanglement witness:
\begin{align}
\mathcal{W}=\frac{1}{2}(4\mathbb{I} -Z_A Z_B -Z_A x_A x_B + X_A z_A X_B + z_A z_B \nonumber \\
 - x_A Z_B x_B - X_A X_B z_B),
\end{align}
where upper case $X$, $Z$ define the Pauli operators for the polarization of the state, lower case $x$, $z$ define the Pauli operators for the momentum. The state is entangled when $-1\leq W< 0$ and particularly it is purely entangled for $\mathcal{W}=-1$. The polarization dependent stabilizers are measured by rotating the analysis waveplates (Fig.~\ref{fig1}). The two beamsplitters perform the transformation in the path of each photon $|d\rangle_i=(|\ell\rangle_i + |r\rangle_i)/\sqrt{2} \to |\ell ' \rangle_i$ and $|a\rangle_i=(|\ell\rangle_i - |r\rangle_i)/\sqrt{2} \to |r ' \rangle_i$, where $|d\rangle_i \leftrightarrow |a\rangle_i$ is achieved by changing the phase $\phi_i$, i.e. tilting the correspondent glass plate on mode $r_i$. This manipulation allows us to measure the momentum-dependent stabilizers involving $x_i$.
We report in Tab.~\ref{tab:stab} the measured outcomes for the stabilizers. The overall value of $\mathcal{W}=-0.634 \pm 0.036$ demonstrates that the state presents genuine multipartite entanglement, and we can derive a lower bound\cite{Toth05} for the fidelity of the created cluster state $F_{|C_4\rangle}\geq \frac{1}{2}(1-\mathcal{W})=0.817 \pm 0.018$. This result is comparable with that of Ref.~\cite{Vallone07}.

\begin{table}[tbh]
\begin{tabular}{|l|l|}
\hline
$Z_A Z_B$     & +0.940 $\pm$ 0.028 \\ \hline
$X_A X_B z_A$ & -0.860 $\pm$ 0.030\\ \hline
$X_A X_B z_B$ & +0.860 $\pm$ 0.030 \\ \hline
$z_A z_B$     & -0.990 $\pm$ 0.007      \\ \hline
$Z_A x_A x_B$ & +0.8092 $\pm$ 0.036 \\ \hline
$Z_B x_A x_B$ & +0.8081 $\pm$ 0.035 \\ \hline
\end{tabular}
\caption{Measured outcomes of the stabilizer used for calculating the genuine multipartite entanglement witness $\mathcal{W}$.}
\label{tab:stab}
\end{table}

\begin{figure}[tbh]
  \centering
\includegraphics[width=0.48\textwidth]{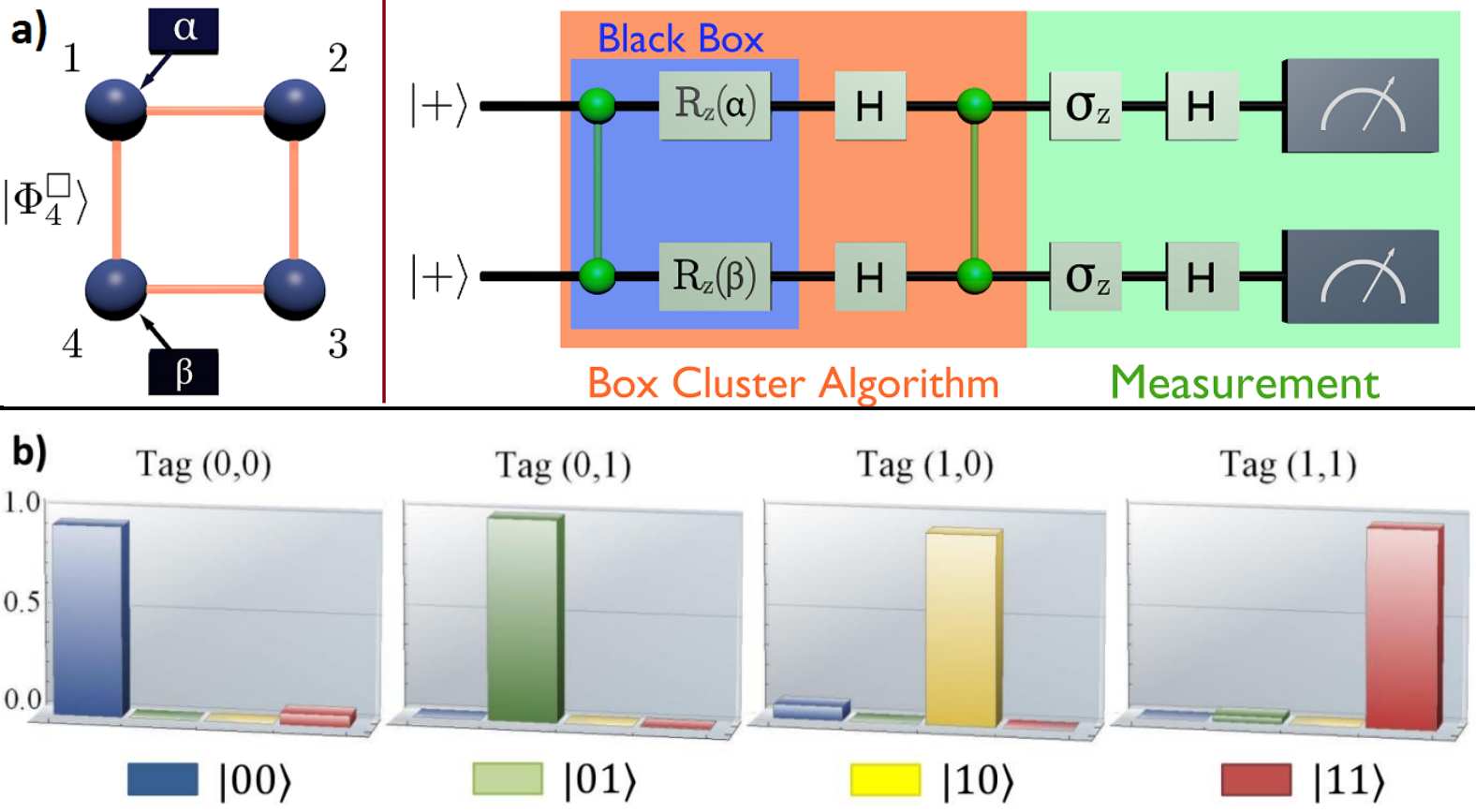}
  \caption{\textbf{Grover's algorithm scheme.} \textbf{a)} Quantum circuit for Grover's algorithm. The black box tags the item through the measurements of qubits 1 and 4. The information is then encoded and processed in  qubits 2 and 3 of the cluster state $|\Phi_4^\Box\rangle$. The single qubit operations are implemented by the choice of the measurement basis. \textbf{b)} Outcome probability for different tagged items for the probabilistic Grover's algorithm. The average success rate of the algorithm is $s=0.960 \pm 0.007$ .
  \label{fig:grover}}
\end{figure}

Finally, the quality of the created cluster state was tested by performing the Grover's search algorithm for a four element database. This is, to the best of our knowledge, the first achievement of a one-way quantum computation basic operation based on multiple DOFs through integrated photonics. The quantum circuit is represented in Fig. \ref{fig:grover}a. It consists of two qubits initially prepared in the state $|+\rangle_1|+\rangle_2$, a black box tagging one item of the database and an operation which allows us to identify the tagged item in the readout. The algorithm can be implemented\cite{Raussendorf01,Browne05} using the four qubit  box cluster state $|\Phi^{\Box}_4\rangle$ defined in Refs.~\cite{Walther05,Vallone08A}. An approach based on a linear optic implementation of the Grover's algorithm using multiple DOFs was originally proposed by Kwiat et al.\cite{Kwiat00}. Following Fig.~\ref{fig:grover}a, the black box tags the item by choosing the bases $\alpha$ and $\beta$ for the measurements on qubits 1 and 4. The information is processed and read on qubits 2 and 3. Labelling the physical qubits in the order $(1,2,3,4)=(k_B,\pi_A,k_A,\pi_B)$, with $k (\pi)$ standing for the path (polarization) qubits, the cluster state $|C_4\rangle$ is equivalent to $|\Phi^{\Box}_4\rangle$ up to the single qubit unitaries
\begin{equation}
\mathcal{U}=\sigma_x H \otimes H \otimes \sigma_z H \otimes H.
\end{equation}
$\mathcal{U}$ can be implemented by simply rotating the measurement basis, as $\mathcal{U}$ is a single qubit transformation.
First, we performed a probabilistic computation where we post-selected the cases with no errors occurring in the one-way computation model\cite{Vallone08A,Vallone08B}. The results are reported in Fig.~\ref{fig:grover}b in which we show that the average success rate in identifying the correct item in the database is $s=0.960 \pm 0.007$ at an average protocol rate of 17~Hz. This result is probabilistic and depends on the postselection of the measurement outcomes. We may then apply a feed-forward protocol in which the outcomes are relabelled depending on the results of the measurements performed by the black box \cite{Walther05,Vallone08A}. In this case the computation is deterministic with a success rate of $s=0.964 \pm 0.003$ and protocol rate of 68~Hz. This procedure of passive feed-forward corresponds to corrections  made in the post-selection process by relabelling the outputs.

\section{Conclusions}

In this work four-qubit hyperentangled states built on the path and polarization of two photons have been adopted to manipulate qubits based on the two degrees of freedom and propagating through an integrated photonic circuit fabricated by the femtosecond laser writing technique.
The device was used to test the presence of path and polarization entanglement. We also engineered cluster states and we measured a multipartite genuine entanglement witness to estimate the purity of correlations within the entangled state. For both experiments we achieved fidelities comparable with previous bulk experiments \cite{Barbieri05,Vallone07}, thus showing the good quality of the experimental results. We exploited the cluster state to perform Grover's search algorithm as an experimental realization on a chip of the one-way quantum computation on chip using different DOFs. Our experiment provides the first demonstration that it is possible to use simultaneously different degrees of freedom of the photons within an integrated photonic circuit, thus increasing the number of qubits. A future step forward in this approach will include the use of more complex circuits for the active manipulation of both phase and polarization inside the same device.

\emph{Note added:} During the review process of this work, the experimental realization of an integrated CNOT gate for measurement-based on-chip quantum computation has been reported \cite{Carolan15}.

\section*{Acknowledgments}
This work was supported by the European Union through the project FP7-ICT-2011-9-600838 (QWAD Quantum Waveguides Application and Development; www.qwad-project.eu) and by FIRB, Futuro in Ricerca HYTEQ.


\begin{widetext}
\centering
SUPPLEMENTARY INFORMATION
\end{widetext}

\section{Path Entanglement}
 \begin{figure}[htbp!]
\centering
\includegraphics[width=0.49\textwidth]{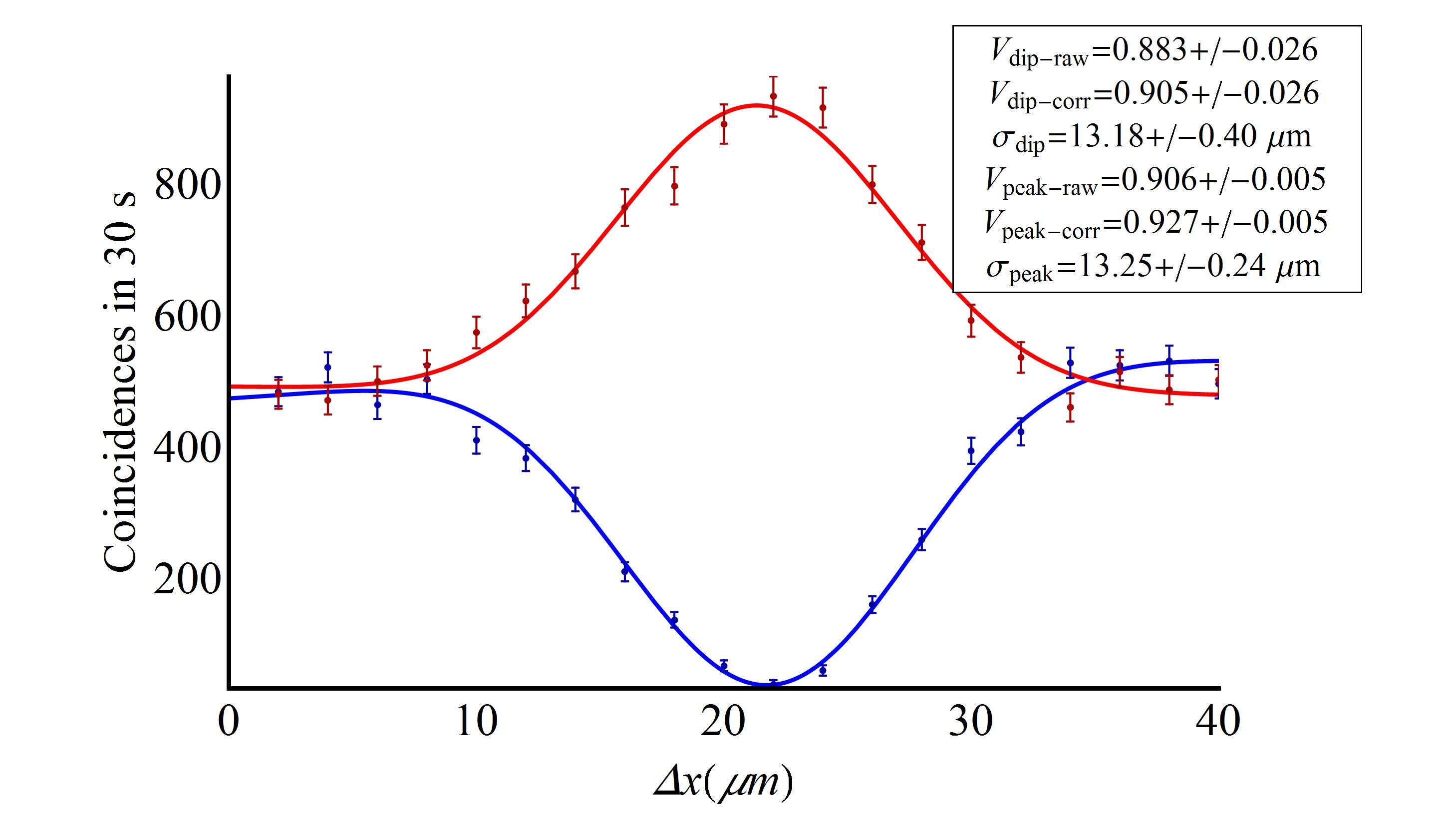}
\caption{Interference pattern between modes $|\ell_A r_B\rangle$ and $|r_A \ell_B\rangle$. Peak is for $\phi=\pi$, dip is for $\phi=0$. Experimental data is fitted with gaussian curves.}
\label{fig:path}
\end{figure}

The presence of path entanglement has been shown by observing interference effects using only one polarization of the light emitted from the source. On this purpose the state $|H_A H_B\rangle \otimes (|r_A \ell_B\rangle + e^{i\phi}|\ell_A r_B\rangle)$ was injected in the integrated device. We observed interference effects between the two couples of modes measuring the visibility of the peak (for $\phi=0$) and the dip (for $\phi=\pi$) in the condition of path indistinguishability. We changed $\phi$ by tilting a glass plate inserted in mode $\ell_B$ and we obtained path indistinguishability varying the length of paths $\ell_B$ and $\ell_A$ using a micrometric translation stage. In Fig. \ref{fig:path} we report as an example the peak and dip as coincidences as function of the delay $\Delta x$. We obtained $V_{peak}=0.892\pm 0.007$ and $V_{dip}=0.915\pm 0.008$ after a fit with a gaussian curve.
 We used 10nm gaussian filters centered in $\lambda=710$nm, the FHWM of the fitted curves have $\sigma=13.2\pm0.02\mu m$.

\section{Polarization Entanglement}

\begin{figure}[htbp!]
  \centering 
  \includegraphics[scale=0.25]{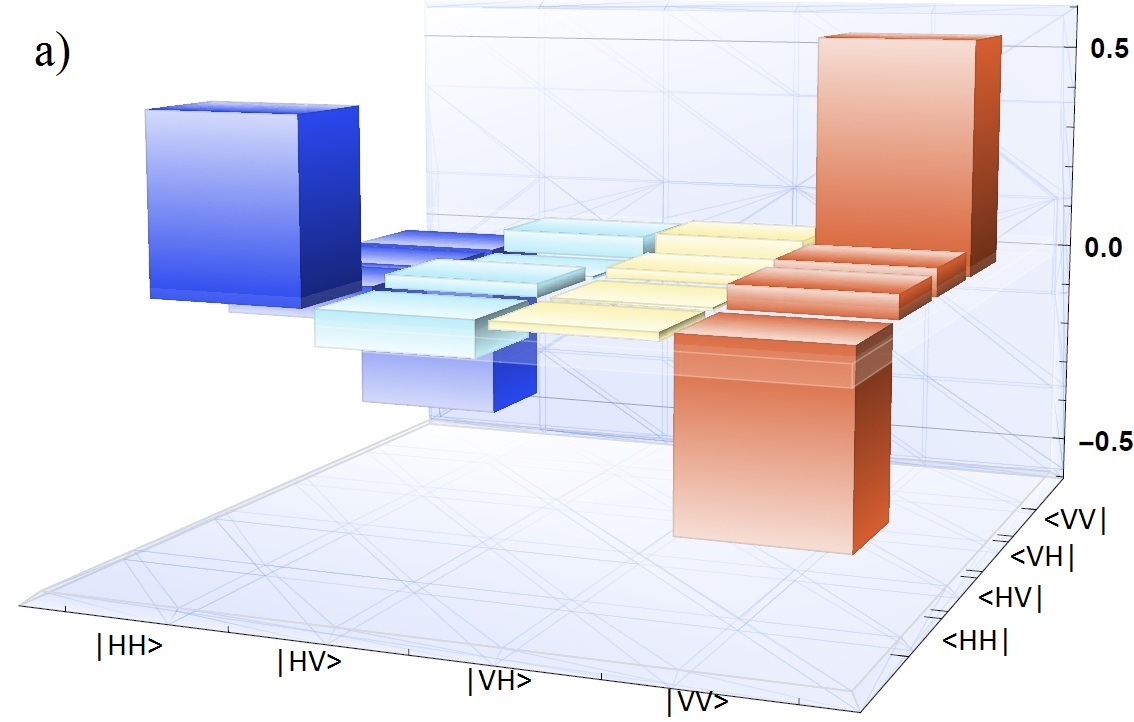}
\includegraphics[scale=0.25]{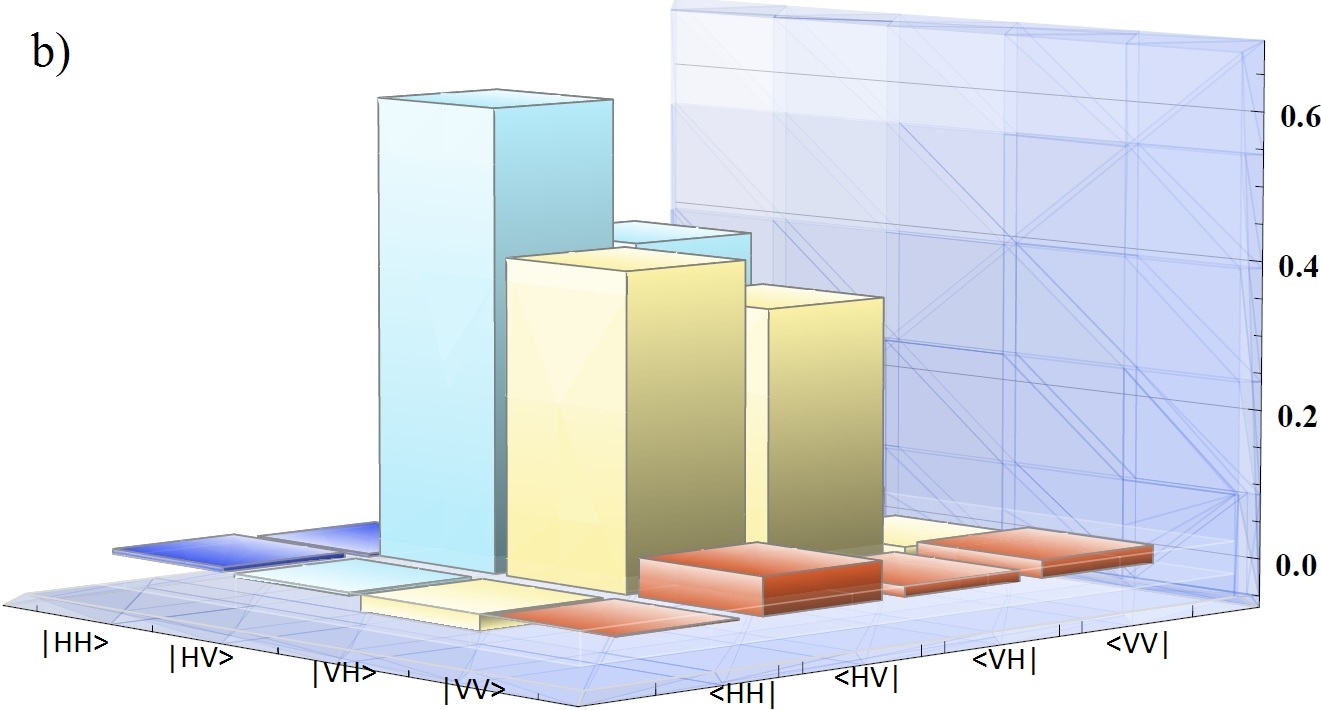}  
\caption{\textbf{a)} Two photon tomography of state $|r_A \ell_B\rangle \otimes |\phi^-\rangle$ performed before the integrated device, (Concurrence$=0.85\pm0.08$, Fidelity$=0.90\pm0.10$, \textbf{b)} Two photon tomography of state $|r_A \ell_B\rangle \otimes |\psi^+\rangle$ performed after polarization compensation and the integrated device (Concurrence$=0.83\pm0.08$, Fidelity$=0.91 \pm 0.09$)) .
  \label{fig:tomo1}}
\end{figure}

We first tested the purity of the source by reconstructing the two qubit-tomography of polarization entangled state $|r_A \ell_B\rangle \otimes \frac{1}{\sqrt{2}}(|H_A H_B\rangle -|V_A V_B\rangle)$  (see Fig \ref{fig:tomo1}a), obtaining a fidelity of $F=0.90\pm0.10$ to the entangled state. Similarly the fidelity of the source to the separable state $|H_A H_B\rangle$ was $F=0.94\pm0.08$. By analyzing a single couple at time we could test polarization entanglement without the contribution given by the path entanglement.

\begin{table}[h!bt]
\begin{tabular}{|l|l|l|l|l|}
\hline
\textbf{Input mode:} & \textbf{Fiber Array} & \textbf{Chip} & \textbf{Collection} & \textbf{Tot} \\ \hline
$\ell_A$               & 59\%                 & 42\%          & 87\%               & 22\%  \\ \hline
$r_A$               & 65\%                 & 41\%          & 87\%               & 23\%       \\ \hline
$\ell_B$               & 48\%                 & 32\%          & 87\%               & 13\%       \\ \hline
$r_B$               & 51\%                 & 40\%          & 87\%               & 18\%       \\ \hline
\end{tabular}
\caption{Measured coupling efficiencies for the four input modes of the fiber array, the integrated device and of the detectors' multimode fibers}
\label{tab:coupling}
\end{table}

In order to test the entanglement in chip we needed to compensate the effect that both the integrated device and the fiber array have on the polarization of the input state. We assumed that these elements perform unitarian transformations to the polarization of the state. As shown in Fig. 1,  we used a QWP (quarter wave plate) and a HWP (half wave plate) to compensate them in the H-V basis, while the phases between $H_i$ and $V_i$ ($i=A,B$) were adjusted tilting on the vertical axis a HWP at $0^\circ$ in the mode $\ell_B$ and using the spherical mirror of the source. We firstly compensated the polarization so that modes $\ell_A$ and $r_A$ behave such that $H \rightarrow H$ and $V \rightarrow V$ while modes $\ell_B$ and $r_B$ behave such that $H \rightarrow V$ and $V \rightarrow H$. Measured fidelity for state $|r_A \ell_B\rangle \otimes \frac{1}{\sqrt{2}}(|H_A V_B\rangle +|V_A H_B\rangle)=|r_A \ell_B\rangle \otimes |\psi^+\rangle$ is F$=0.91\pm0.09$ (see Fig. \ref{fig:tomo1}b), thus showing that the integrated devices preserve the polarization entanglement.

\end{document}